\documentclass[onecolumn,showpacs,nofootinbib,aps,superscriptaddress, eqsecnum,prd,notitlepage,showkeys,10pt]{revtex4-1}
\usepackage{amssymb}
\usepackage{amsmath}
\usepackage{graphicx}
\usepackage{subfigure}
\usepackage{dcolumn}
\usepackage{natbib}
\usepackage{hyperref}
\newcommand{\ben}{\begin{eqnarray}}
\newcommand{\een}{\end{eqnarray}}
\newcommand{\be}{\begin{equation}}
\newcommand{\ee}{\end{equation}}
\newcommand{\n}{\label}
\newcommand{\no}{\noindent}

\begin{document}

\title{The extended holographic Dark Energy cosmological models}

\author{M\'onica Forte}
\email{forte.monica@gmail.com}
\affiliation{Departamento de F\'isica, Facultad de ciencias Exactas y Naturales, Universidad de Buenos Aires, 1428 Buenos Aires, Argentina}

\begin{abstract}
We present a general analytical treatment of cosmological models commanded by three interactive fluids with arbitrary barotropic indexes. The variable equations of state case are applicable to universes with quintessences, k essences and holographic fluids and we pay some attention to the holographic case. We also propose the need to extend the functional forms of the holographic energy densities that are usual in the literature  in the face of the incompatibility presented by them when special interactions are used.

\end{abstract}
\maketitle

\section{Introduction}


The enormous volume of observational astronomical data \cite{Supernova,SDSS,Spergel:2003cb,Dark,Wetterich:1987fmyRatra:1987rm,Peebles:2002gy,Kolb:2005da} 
shows us a universe well adjusted by the model $\Lambda$CDM with respect to flatness, expansion and acceleration. However, there are controversies that current physics tries to settle in different ways. One widely used way is to assume that there are interactions in the dark sector \cite{Ellis_Wetterich_Amendola_Gasperini,interacc,Forte:2015oma,Chen:2011cy,Bertolami:2007zm,Bertolami:2007tq,Abdalla:2007rd,He:2009pd,Guo:2007zk,Quartin:2008px,Pavon:2007gt} that explain, for example, the difference in energy density values between the tentative calculations assigned to the primordial vacuum and what is currently measured as dark energy. Other discrepancies refer to the coincidence between ranges of magnitude of the components of dark matter and dark energy and also to the unexplained evidence that assigns greater longevity to certain some old high redshift objects (OHROs), old galaxy and quasar discovered, compared to what is considered the age of our universe \cite{Forte:2013fua}. 
These disagreements have impelled to consider interactive models that generally only contain two components. In particular, two-component models have been developed where a holographic fluid plays the role of dark energy 
\cite{Nojiri:2005pu,Li:2004rb,Wang:2005jx,Forte:2016ben,Bamba:2012cp} and has been shown to be intrinsically interactive \cite{Chimento:2011dw}.The next step has been to consider cosmological models of three interacting components with 
constant barotropic indices \cite{Chimento:2013se,Chimento:2013ira}.
In this work, we present an analytical treatment of cosmological models commanded by three interactive fluids with arbitrary barotropic indexes, constant \cite{Forte:2018loa} or not. This last case is applicable to universes with quintessences, k-essences and holographic fluids. We also propose the need to extend the functional forms of the holographic energy densities that are usual in the literature  in the face of the incompatibility presented by them when special interactions are used.

The paper is organized as follows. In Sec. II the continuity equations for each interacting fluid are presented in terms of constant auxiliary real parameters and modified interactions. They are defined there and are included in the master differential equation whose resolution yields the functional form of the global energy density. The functional forms for the energy densities of each of the individual fluids are also written in terms of those auxiliary constants, the global energy density and their derivatives and the modified interactions.
In Sec. III this methodology is applied to the case in which the barotropic indexes of all the interacting fluids are constant, identifying the auxiliary parameters with these indexes and analyzing particular cases.
In Sec. IV  this methodology is applied to the case in which at least one of the barotropic indices is variable and the restrictions on the values to be assigned to these parameters are shown. Also, we analyze particular cases for which the interactive part of the master equation is canceled.
In Sec. V we study cases of interactive holographic fluids with special interactions that show the incompatibility of the holographic models most used in the literature for three fluid systems and we propose a new functional form that includes a term proportional to the jerk.
Finally in section VI we present the conclusions.

\section{Three interacting fluids with general equations of state  }
\label{section2}

In Friedmann Lema\^{i}tre Robertson Walker (FLRW) background geometry we consider a cosmological model commanded by three interactive perfect fluids with pressures and densities of energy  $p_i,\rho_i$,  $i=1,2,3$ and variable barotropic indexes  $\gamma_i=1+p_i / \rho_i$ that satisfy conservation equations 
\ben
\n{1}
\dot\rho_1 + 3H\gamma_1\rho_1= 3H\mathbb{Q}_1,  \\
\n{2}
\dot\rho_2 + 3H\gamma_2\rho_2= 3H\mathbb{Q}_2,  \\
\n{3}
\dot\rho_3 + 3H\gamma_3\rho_3= 3H\mathbb{Q}_3.
\een
\no where the logarithmic derivative of the factor of scale $a$ with respect to cosmological time $H=\dot a/a$ is the Hubble parameter and $\mathbb{Q}_i$ are the interactions affecting the fluids that satisfy the condition $\sum_{i=1}^3 \mathbb{Q}_i=0$. 

  It is convenient to write the partial conservation equations for each fluid in terms of the constant auxiliary parameters $\alpha_i$ that facilitate the resolution of the equations. Then

\ben
\n{4}
\rho_1' + \alpha_1\rho_1=Q_1,  \\
\n{5}
\rho_2' + \alpha_2\rho_2=Q_2,  \\
\n{6}
\rho_3' +\alpha_3\rho_3=Q_3,
\een

\no where  the $'$ imply derivative with respect to the variable $\eta=\ln a^3$ and each $Q_i$ is defined as $Q_i=\mathbb{Q}_i+(\alpha_i - \gamma_i)\rho_i $.
The Einstein equations for this model are
\ben
\n{7}
3H^2& = &\rho,  \\
\n{8}
2\dot H & = & - \gamma\rho = - \gamma_1\rho_1 - \gamma_2\rho_2 - \gamma_3\rho_3 = \rho'.
\een

Using (\ref{4}), (\ref{5}), (\ref{6}) and the equation obtained by derivation of (\ref{8}) we can write the system of equations,
\ben
\n{9}
\rho & = & \sum_{i=1}^3\rho_i, \\
\n{10}
\rho' - \sum_{i=1}^3 Q_i & = & - \sum_{i=1}^3 \alpha_i \rho_i,\\
\n{11}
\rho'' + \mathbf{\alpha} \cdot \mathbf{Q} - \sum_{i=1}^3 Q'_i& = &\sum_{i=1}^3\alpha_i ^2\rho_i,
\een
\no in order to describe the densities of energy of each fluid in terms of the total energy density and its first and second derivatives as 

\ben
\n{12}
\frac{\Delta}{\Delta_{23}}\rho_1= \rho''+(\alpha_2+\alpha_3)\rho'+\alpha_2\alpha_3\rho+\mathbf{\alpha} \cdot \mathbf{Q} - \sum_{i=1}^3Q'_i -(\alpha_2+\alpha_3)\sum_{i=1}^3 Q_i, \\
\n{13}
-\frac{\Delta}{\Delta_{13}}\rho_2= \rho''+(\alpha_1+\alpha_3)\rho'+\alpha_1\alpha_3\rho+ \mathbf{\alpha} \cdot \mathbf{Q} - \sum_{i=1}^3 Q'_i -(\alpha_1+\alpha_3)\sum_{i=1}^3 Q_i, \\
\n{14}
\frac{\Delta}{\Delta_{12}}\rho_3= \rho''+(\alpha_1+\alpha_2)\rho'+\alpha_1\alpha_2\rho+\mathbf{\alpha} \cdot \mathbf{Q} - \sum_{i=1}^3 Q'_i -(\alpha_1+\alpha_2)\sum_{i=1}^3Q_i.
\een

\no Above  $\Delta$ is the determinant of the system of equations formed by (\ref{9}), (\ref{10}) and (\ref{11}), that is, $\Delta=\Delta_{12}\Delta_{13}\Delta_{23}$ with $\Delta_{ij}=(\alpha_i-\alpha_j)$ and the expression $\mathbf{\alpha} \cdot \mathbf{Q}$ means $\mathbf{\alpha} \cdot \mathbf{Q}=\alpha_1Q_1+ \alpha_2Q_2+\alpha_3Q_3$.

At this point we can obtain the equation which resolution bring us the functional form of the global density of energy $\rho$. When we differentiate the equation (\ref{12}) and then apply to it the equations (\ref{4}) and (\ref{12}), we obtain a result similar to the obtained when we differentiate the equation (\ref{13}) and then apply to it the equations (\ref{5}) and (\ref{13}) or to the obtained when we differentiate the equation (\ref{14}) and then apply to it the equations (\ref{6}) and (\ref{14}). The sum of these three similar results allows to obtain the expression 
\be
\begin{split}
\n{15}
\rho''' + (\alpha_1+\alpha_2+\alpha_3)\rho'' + (\alpha_1\alpha_2+\alpha_1\alpha_3 +\alpha_2\alpha_3)\rho' &+ \alpha_1\alpha_2\alpha_3 \rho =  \\
\sum_{i=1}^3 Q_i''  + (\sum_{i=1}^3 \alpha_i )(\sum_{i=1}^3 Q_i')  -  \mathbf{\alpha} \cdot \mathbf{Q}' + Q_1\alpha_2\alpha_3 & + Q_2\alpha_1\alpha_3 + Q_3\alpha_1\alpha_2.
\end{split}
\ee

The equation (\ref {15}) can be called the master equation for cosmological models filled with three interactive perfect fluids.


\section{The case with constant barotropic indexes}


The values considered for the constants $ \alpha_i $ depend on the particular model. When all the barotropic indexes $ \gamma_i $ are constant, a suitable choice is $\alpha_i = \gamma_i $ for all i with which (\ref {15}) is reduced to
\be
\begin{split}
\n{16}
\rho'''+(\gamma_1+\gamma_2+\gamma_3)\rho'' + (\gamma_1\gamma_2+\gamma_1\gamma_3+\gamma_2\gamma_3)\rho' + \gamma_1\gamma_2\gamma_3 \rho & =  \\
 -\mathbf{\gamma} \cdot\mathbb{Q}' +\mathbb{Q}_1 \gamma_2\gamma_3 + \mathbb{Q}_2\gamma_1\gamma_3 &+ \mathbb{Q}_3\gamma_1\gamma_2.
\end{split}
\ee

For the natural options for constant indices, $\alpha_i = \gamma_i $, and due to the condition of continuity, $\sum_{i=1}^3\mathbb{Q}_i=0$, there are different cases for which each interaction $\mathbb{Q}_i$ is written as a linear function of a single arbitrary interaction $\mathcal{Q}$.\\ 
The so-called transverse case $\mathbf {\gamma}\cdot\mathbb {Q} = 0 $ used in \cite{Chimento:2012aea}, cancels the term $ \mathbf {\gamma} \cdot\mathbb {Q}' = 0 $ and leads to write, for example, $\mathbb{Q}_1=\mathcal{Q}$,  $\mathbb{Q}_2=\Delta_{31}\mathcal{Q}/\Delta_{23}$ and $\mathbb{Q}_3=\Delta_{12}\mathcal{Q}/\Delta_{23}$. Here, (\ref {16}) is written as 
\be
\n{16b}
\rho'''+(\gamma_1+\gamma_2+\gamma_3)\rho'' + (\gamma_1\gamma_2+\gamma_1\gamma_3+\gamma_2\gamma_3)\rho' + \gamma_1\gamma_2\gamma_3 \rho = \gamma_1\psi\mathcal{Q},\qquad \psi=\frac{\gamma_1\gamma_2\Delta_{12}+\gamma_3\gamma_1\Delta_{31}+\gamma_2\gamma_3\Delta_{23}}{\gamma_1\Delta_{23}}.
\ee

In the non-transverse case $\mathbf{\gamma}\cdot\mathbb{Q}\not= 0, $ the condition $\mathbb{Q}_1 \gamma_2\gamma_3 + \mathbb{Q}_2\gamma_1\gamma_3 + \mathbb{Q}_3\gamma_1\gamma_2=0$, allows to write  $\mathbb{Q}_1=\mathcal{Q}$, $\mathbb{Q}_2= (\gamma_2/\gamma_1)\Delta_{31} \mathcal{Q}/\Delta_{23}$ and  $\mathbb{Q}_3= (\gamma_3/\gamma_1)\Delta_{12} \mathcal{Q}/\Delta_{23}$ and then (\ref {16})  is written as 
\be
\n{16c}
\rho'''+(\gamma_1+\gamma_2+\gamma_3)\rho'' + (\gamma_1\gamma_2+\gamma_1\gamma_3+\gamma_2\gamma_3)\rho' + \gamma_1\gamma_2\gamma_3 \rho = -\psi\mathcal{Q}'.
\ee

When the non-transverse set of interactions satisfies the condition, $ -\mathbf{\gamma} \cdot\mathbb{Q}' +\mathbb{Q}_1 \gamma_2\gamma_3 + \mathbb{Q}_2\gamma_1\gamma_3 + \mathbb{Q}_3\gamma_1\gamma_2=0$, equivalent to the relationship $\Delta_{12}(\mathbb{Q}_2'+\gamma_3\mathbb{Q}_2)=\Delta_{31}(\mathbb{Q}_3'+\gamma_2\mathbb{Q}_3)$, it is possible to consider the case in which both numerator and denominator are constants. 
In this supposition, the functional forms of the $\mathbb{Q}_i$ are
\ben
\n{17a}
\mathbb{Q}_1&=&\mathbb{A}_1\frac{\gamma_1\Delta_{23}}{\gamma_2\gamma_3\Delta_{12}}-\frac{\mathbb{A}_2}{ a^{3\gamma_2}} -\frac{\mathbb{A}_3}{a^{3\gamma_3}},  \\
\n{17b}
\mathbb{Q}_2&=&\mathbb{A}_1\frac{\Delta_{31}}{\gamma_3\Delta_{12}} + \frac{\mathbb{A}_3}{a^{3\gamma_3}},  \\
\n{17c}
\mathbb{Q}_3&=&\frac{ \mathbb{A}_1}{\gamma_2}+\frac{\mathbb{A}_2}{a^{3\gamma_2}},
\een
\no with $\mathbb{A}_i$ constants of integration. 

A specific case with a non-transverse set of interactions is discussed in \cite{Forte:2018loa} in an approach to the cosmological constant problem. There, the $\mathbb{Q}_i$ interactions $\mathbb{Q}_1=\mu(\rho-\rho'')$ and $\mathbb{Q}_2=\alpha \rho'$, do not satisfy any of the above conditions, neither $ \mathbf {\gamma} \cdot\mathbb {Q}' = 0 $ nor $\mathbb{Q}_1 \gamma_2\gamma_3 + \mathbb{Q}_2\gamma_1\gamma_3 + \mathbb{Q}_3\gamma_1\gamma_2=0$ nor $ -\mathbf{\gamma} \cdot\mathbb{Q}' +\mathbb{Q}_1 \gamma_2\gamma_3 + \mathbb{Q}_2\gamma_1\gamma_3 + \mathbb{Q}_3\gamma_1\gamma_2=0$.


\section{The case with variable barotropic indexes}


We now consider that at least one of the barotropic indices is variable. 
This is an interesting situation because it allows to include fluids like quintessences, k-essences and holographics in this approach of models with three interactive fluids. 
The simplest solutions of the eq. (\ref{15}) correspond to the systems where the interactions satisfy the condition 
\be
\n{17}
\sum_{i=1}^3 Q_i''  + (\sum_{i=1}^3 \alpha_i )(\sum_{i=1}^3 Q_i')  -  \mathbf{\alpha} \cdot \mathbf{Q}' + Q_1\alpha_2\alpha_3  + Q_2\alpha_1\alpha_3 + Q_3\alpha_1\alpha_2=0.
\ee
The trivial solution of  (\ref {17}) where each $Q_i=0$, allows us to draw very interesting conclusions. These modified interactions correspond to having bare interactions $\mathbb{Q}_i$ on each of the three fluids of the form $\mathbb{Q}_i=\gamma_i\rho_i-\alpha_i\rho_i$. Then each density of energy  results of the form $\rho_i= \rho_{io}a^{-3\alpha_i}$ as can be seen from (\ref {12}), (\ref {13}), (\ref {14}) once solved (\ref {15}), or directly from the equations of continuity  (\ref {4}), (\ref {5}), (\ref {6}). The equivalent model corresponds to a system with three self-preserved fluids and this happens in spite of the fact that the barotropic indices are variable. This result appears to be trivial but entails an important restriction on the form of any holographic fluid when this option can be considered as the DE component of the system. It shows the need to use more general holographic fluids than those seen in the literature until now.

A barely more general solution of (\ref {17}) where it is still true that the sum is null, $\sum_{i=1}^3 Q_i=0$, in spite of the fact that each term is not canceled separately, results in the modified continuity equation
\be
\n{18}
-\rho'= \sum_{i=1}^3 \gamma_i \rho_i = \sum_{i=1}^3 \alpha_i \rho_i,
\ee
 This result is similar to that obtained in two fluids systems, where one of them is a modified holographic type Ricci fluid acting as dark energy \cite{Chimento:2013se}, but here it is obtained in a very different form and it is fundamentally due to the set of interactions chosen, regardless of the type of fluids considered since the global energy conservation prescribes $\sum_{i=1}^3 \mathbb{Q}_i = 0$. 

When only one fluid has variable barotropic index, the condition $\sum_{i=1}^3 Q_i = 0$ leads to restrictions on the type of interactions and on the choice of the constant auxiliary coefficients $\alpha_i$. Trivially, if for example  $Q_1 \not= 0$, then only can be $Q_2=0$ or $Q_3=0$  but not both. With respect to the choice of constants, if for example $\gamma_2$ is the only one variable index, then it can be chosen $\alpha_1=\gamma_1$ or $\alpha_3=\gamma_3$ but not both, because the index $\gamma_2$ is a linear combination with constant coefficients of the quotients $\rho_1/\rho_2$ and $\rho_3/\rho_2$, $\gamma_2 = \alpha_2 + (\alpha_1-\gamma_1)\rho_1/\rho_2+(\alpha_3-\gamma_3)\rho_3/\rho_2$.
When, in addition, it is verified that the modified interactions $ Q_i$ satisfy the condition 
\be
\n{19}
(\mathbf{\alpha} \cdot \mathbf{Q})'=  Q_1\alpha_2\alpha_3+Q_2\alpha_3\alpha_1+Q_3\alpha_1\alpha_2,
\ee
\no the problem results in a global density of energy that looks like a system with three self-preserved fluids $\rho=3H_0^2 \sum_{i=1}^3 b_i/a^{3\alpha_i}$ with $\sum_{i=1}^3 b_i=1$. 

However now, each fluid is an interacting fluid with densities of energy 
\ben
\n{20a}
\Delta_{12}\Delta_{13}\rho_1 = 3H_0^2 \sum_{i=1}^3\frac{b_i\Delta_{i2}\Delta_{i3}}{a^{3\alpha_i}} + \mathbf{\alpha} \cdot \mathbf{Q},  \\
\n{20b}
\Delta_{12}\Delta_{23}\rho_2 = 3H_0^2 \sum_{i=1}^3\frac{b_i\Delta_{1i}\Delta_{i3}}{a^{3\alpha_i}} + \mathbf{\alpha} \cdot \mathbf{Q},  \\
\n{20c}
\Delta_{13}\Delta_{23}\rho_3 = 3H_0^2 \sum_{i=1}^3\frac{b_i\Delta_{1i}\Delta_{2i}}{a^{3\alpha_i}} + \mathbf{\alpha} \cdot \mathbf{Q},
\een
Then, (\ref{19}) can be rewritten as 
\be
\n{20d}
\Delta_{31}[Q_3' +\alpha_2 Q_3]= \Delta_{12}[Q_2' +\alpha_3 Q_2], 
\ee
\no as in the constant case and admits solutions like (\ref{17a}), (\ref{17b}) and (\ref{17c}), but with the corresponding replacements $\gamma_i \rightarrow \alpha_i$.

In the general case, when the interactions are linear functions of the global density $\rho$ and its derivative $\rho'$,
\be
\n{210} Q_i = a_{i0} + a_{i1}\rho + a_{i2}\rho',\qquad \i=1,2,3,
\ee
\no the equation (\ref {15}) is written as
\be
\n{21}
\mathcal{A}_0 + \mathcal{A}_1 \rho + \mathcal{A}_2\rho' + \mathcal{A}_3\rho'' + \mathcal{A}_4\rho''' = 0,
\ee
\no with 

\begin{eqnarray*}
\mathcal{A}_0 & = & -[a_{10}\alpha_2\alpha_3+ a_{20}\alpha_1\alpha_3+ a_{30}\alpha_1\alpha_2], \\
\mathcal{A}_1 & = & -[a_{11}\alpha_2\alpha_3+ a_{21}\alpha_1\alpha_3+ a_{31}\alpha_1\alpha_2]+\alpha_1\alpha_2\alpha_3, \\
\mathcal{A}_2 & = & \alpha_1\alpha_2(1-a_{32})+\alpha_1\alpha_3(1-a_{22})+\alpha_2\alpha_3(1-a_{12})-\sum_{i=1}^3\alpha_i\sum_{i=1}^3 a_{i1}+\sum_{i=1}^3\alpha_ia_{i1}, \\
\mathcal{A}_3 & = & \sum_{i=1}^3\alpha_i-\sum_{i=1}^3a_{i1}-\sum_{i=1}^3\alpha_i\sum_{i=1}^3a_{i2}+\sum_{i=1}^3\alpha_ia_{i2}, \\
\mathcal{A}_4 & = & 1-\sum_{i=1}^3a_{i2}.                    
\end{eqnarray*}

The general solution of equation (\ref {21}) is greatly simplified with an adequate choice of auxiliary parameters $\alpha_i$. For example we can choose them in such a way that $\mathcal{A}_0=\mathcal{A}_1=0$. That is 

\begin{eqnarray*}
\alpha_1 & = & \frac{[a_{11}a_{30}-a_{10}a_{31}]}{[a_{20}a_{31}-a_{21}a_{30}+\alpha_2a_{30}]}\alpha_2, \\
\alpha_3 & = & \frac{a_{30}a_{11}-a_{10}a_{31}}{[a_{10}a_{21}-a_{11}a_{20}-a_{10}\alpha_2]}\alpha_2.           
\end{eqnarray*}

With these simplifications, the global density of energy turns out to be
\be
\n{22}
\rho=3H_0^2\big(c_0+\frac{c_+}{a^{3\lambda_+}}+\frac{c_+}{a^{3\lambda_+}}\big)\qquad c_0+c_++c_- = 1,
\ee
\no and the remaining parameter $\alpha_2$ can be adjusted observationally just like the actual Hubble parameter $H_0$. The additional condition on linear interactions (\ref {210}) $\sum_{i=1}^3a_{i2}=1$ ensures that the roots $\lambda_{\pm}$ are always real
\begin{equation*}
\lambda_{\pm}= \frac{\mathcal{A}_3\pm\sqrt{\mathcal{A}_3^2-4\mathcal{A}_4\mathcal{A}_2}}{2\mathcal{A}_4}.
\end{equation*}

 
\section{The holographic type dark energy for systems with three components}


From here on we will apply the above considerations to systems with two fluids of constant indices $\gamma_1$ and $\gamma_3$ plus a holographic fluid of variable barotropic index $\gamma_2$.  In general, the characteristic length in the cosmological application of holographic fluids has been considered proportional to $H^{-1}\sim \rho^{-1/2}$, or $\dot H^{-1/2}\sim {\rho'}^{-1/2}$, that is, the density of dark energy is proportional to $H^2\sim \rho$, or $\dot H\sim \rho'$ or their linear combinations. 
As result of that, the holographic density of energy has been taken in the form $\rho_{holo}\sim (g_0H^2+g_1\dot H)\sim  (g_2\rho+g_3\rho') $, with $g_i$ different constants of proportionality. In these cases there  will be compatibility with the interactive models only  when modified interactions meet the condition 
\be
\n{23}
\rho''+ \mathbf{\alpha} \cdot \mathbf{Q} - \sum_{i=1}^3 Q'_i -(\alpha_1+\alpha_3)\sum_{i=1}^3 Q_i=0.
\ee
Note that this incompatibility occurs not only in the very particular case of each $Q_i=0$ but also in cases of transverse interactions $\mathbf{\alpha} \cdot \mathbf{Q}=0$ with $\sum_{i=1}^3 Q_i =0$. If the condition (\ref{23}) is not satisfied, the equation (\ref {13}) shows that useful holographic forms of density of energy $\rho_{holo}\sim g_2\rho+g_3\rho' $ are incompatible with three-component systems. This incompatibility generates the necessity of extending the forms used up to now including some term proportional to jerk $j=-[1+9(\rho'+\rho'')/2\rho]$ that contributes a term with $\rho''$\cite{Dunajski:2008tg,Wang:2005jx,Kim:2005at}.

The existence of a non-accelerated cosmological period followed by an accelerated era implies that the jerk must be non-zero and in fact, the most accepted cosmological model $\Lambda$CDM model itself has $j=1$. Note, that for the compatibility issue of the widely used holographic Granda's form,  it is enough that the jerk is constant, because in that case the second derivative $\rho''$ is simply a linear combination of $\rho$ and $\rho'$.
With this characteristic in mind it is acceptable to propose a new form of holographic energy density that adds a new term proportional to $\rho''$ and that occurs naturally in the definition of the jerk.

Let us study a simple model that includes the extended holographic fluid to show how to choose the constants $\alpha_i$ and get results that can be adjusted observationally. Be $\gamma_1$ and $\gamma_3$ the constant barotropic indices of two perfect fluids with density of energy $\rho_1$ and $\rho_3$ respectively, that interact with an extended holographic fluid with density of energy
\be
\n{24}
\rho_2^{ext}=\rho\Big(\nu_0-\nu_1\gamma-2\nu_2j\Big)=(\nu_0+2\nu_2)\rho+(\nu_1+9\nu_2)\rho'+9\nu_2\rho'',
\ee
\no through a transverse set of interactions $\mathbf{\alpha} \cdot \mathbf{Q}=0$ such that $\sum_{i=1}^3 Q_i =0$. Note that the expression (\ref{24}) are only an extended version of the modified holographic Ricci type density of energy used in \cite{Chimento:2011dw},\cite{Granda:2008dk}. Comparing (\ref{13}) with (\ref{24}) and under the condition $\nu_1(\nu_1+18\nu_2)+9\nu_2(\nu_2-4\nu_0)\geq 0$, the constants $\alpha_i$ must have the values 

\begin{align}
\label{25}
\begin{split}
\alpha_1&=\frac{\nu_1+9\nu_2\pm \sqrt{(\nu_1+9\nu_2)^2-36\nu_2(\nu_0+2\nu_2)}}{18\nu_2},  \\
\alpha_2&=\frac{\nu_1+9\nu_2\pm \sqrt{(\nu_1+9\nu_2)^2-36\nu_2(\nu_0+2\nu_2-1)}}{18\nu_2},  \\
\alpha_3&=\frac{\nu_0+2\nu_2}{(\nu_1+9\nu_2)\alpha_1-(\nu_0+2\nu_2)}\alpha_1.
\end{split}
\end{align}

The transverse set of modified interactions with zero sum allows expressing the three interactions as functions of a single one, say $\mathcal{Q}$, as 
\be
\n{26}
Q_1=\frac{\Delta_{23}}{\Delta_{31}}\mathcal{Q}, \quad Q_2=\mathcal{Q}, \quad Q_3=\frac{\Delta_{12}}{\Delta_{31}}\mathcal{Q}.
\ee
So, if $\mathcal{Q}$ looks like $\mathcal{Q}= \frac{\Delta_{31}}{R}\Big(\rho'''+ G[\rho_1,\rho_2,\rho_3,\rho,\rho',\rho'']\Big)$ with $R=\alpha_1\alpha_2\Delta_{12}+\alpha_2\alpha_3\Delta_{23}+\alpha_3\alpha_1\Delta_{31}$ and $G[\rho_1,\rho_2,\rho_3,\rho,\rho',\rho'']$  an arbitrary function of the densities of energy $\rho_1$, $\rho_2$, $\rho_3$, $\rho$,$\rho'$ and $\rho''$, the equation (\ref{15}) takes the form

\be
\n{27}
(\alpha_1+\alpha_2+\alpha_3)\rho'' + (\alpha_1\alpha_2+\alpha_1\alpha_3 +\alpha_2\alpha_3)\rho'+ \alpha_1\alpha_2\alpha_3 \rho = G[\rho_1,\rho_2,\rho_3,\rho,\rho',\rho''].
\ee
Its solutions have already been obtained for many functions G that are linear or non-linear combinations of its arguments in \cite{Chimento:2009hj}.

In any of the transverse cases with zero sum, $\sum_{i=1}^3 Q_i=0$, the global conservation equation (\ref{18}) allows writing the variable equation of state for the extended holographic dark energy (\ref{24}) $\omega_2^{ext}=\gamma_2-1$ as 

\be
\n{28}
\omega_2^{ext} = \alpha_2-1 + \frac{\alpha_3\rho'' + \alpha_2\rho' + \alpha_1\rho}{\big(\rho'' + (\alpha_1+\alpha_3)\rho' + \alpha_1\alpha_3\rho\big)(\alpha_1-\alpha_3)},
\ee

\no where the auxiliary parameters $\alpha_i$ are given by equations (\ref{25}), the barotropic indices $\gamma_1$ and  $\gamma_3$ can be variable or not, and $\rho$, $\rho'$, $\rho''$ are obtained by resolving (\ref{27}).


\section{Conclusions}


In this work, the equations of evolution for cosmological systems composed of three perfect fluids in interaction are presented. The novelty with respect to other similar presentations is that the expressions contain constant auxiliary parameters $\alpha_i$ that allow to include fluids with constant and non-constant barotropic indices such as quintessence, k essences and holographic fluids. For these last cases we show the incompatibilities derived from using holographic fluids with energy densities that only include terms proportional to the global density of energy $\rho$ and/or to its first temporal derivative $\rho'$.
When all the barotropic indices are constant, the most practical option is to assimilate each auxiliary parameter with each constant barotropic index that is, $\alpha_i=\gamma_i$. When at least one of the barotropic indices is variable, solutions must be found for the general master equation and the choice of values for each parameter depends strongly on the particular problem. If we apply this approach to the case of holographic fluids, it is observed that in some cases there are incompatibilities for the use of the usual forms reported in the literature, presenting the need for an extended holographic fluid whose energy density includes a term proportional to the jerk. The use of an extended holographic fluid as dark energy completely determines the values of the auxiliary parameters and  for these cosmological models we present a study of the cases corresponding to sets of interactions, transverse or not, whose sum is null. The expressions of all the densities of energy involved are described as well as the shape adopted by the equation of state of the holographic dark energy density in the case of modified interactions whose sum is zero. Another important characteristic of this type of approach is that the constant barotropic indexes of the non-holographic fluids that interact with the holographic, extended or not, do not need to be different and that is why they include the cosmological models that do require it as in \cite{Chimento:2013qja}.



\begin{thebibliography}{99}

\bibitem{Supernova} 
  A.~G.~Riess {\it et al.}  [Supernova Search Team Collaboration],
  Astron.\ J.\  {\bf 116}, 1009 (1998)
  [astro-ph/9805201];
S.~Perlmutter {\it et al.}  [Supernova Cosmology Project Collaboration],
  Astrophys.\ J.\  {\bf 517}, 565 (1999)
  [astro-ph/9812133].
\bibitem{SDSS} 
  M.~Tegmark {\it et al.}  [SDSS Collaboration],
  Phys.\ Rev.\ D {\bf 69}, 103501 (2004); 
  K.~Abazajian {\it et al.}  [SDSS Collaboration],
  Astron.\ J.\  {\bf 128}, 502 (2004)
  [astro-ph/0403325];
  K.~Abazajian {\it et al.}  [SDSS Collaboration],
  Astron.\ J.\  {\bf 129}, 1755 (2005)
  [astro-ph/0410239].
\bibitem{Spergel:2003cb} 
  D.~N.~Spergel {\it et al.}  [WMAP Collaboration],
  Astrophys.\ J.\ Suppl.\  {\bf 148}, 175 (2003)
  [astro-ph/0302209].
\bibitem{Dark} 
  V.~Sahni and A.~A.~Starobinsky,
  Int.\ J.\ Mod.\ Phys.\ D {\bf 9}, 373 (2000)
  [astro-ph/9904398];
  P.~J.~E.~Peebles and B.~Ratra,
  Rev.\ Mod.\ Phys.\  {\bf 75}, 559 (2003)
  [astro-ph/0207347];
  V.~Sahni,
  Lect.\ Notes Phys.\  {\bf 653}, 141 (2004)
  [astro-ph/0403324];
  E.~J.~Copeland, M.~Sami and S.~Tsujikawa,
  Int.\ J.\ Mod.\ Phys.\ D {\bf 15}, 1753 (2006)
  [hep-th/0603057];
  J.~Frieman, M.~Turner and D.~Huterer,
  Ann.\ Rev.\ Astron.\ Astrophys.\  {\bf 46}, 385 (2008)
  [arXiv:0803.0982 [astro-ph]];  S.~'i.~Nojiri and S.~D.~Odintsov,
  Phys.\ Rept.\  {\bf 505}, 59 (2011)
  [arXiv:1011.0544 [gr-qc]];
  M.~Li, X.~-D.~Li, S.~Wang and Y.~Wang,
  Commun.\ Theor.\ Phys.\  {\bf 56}, 525 (2011)
  [arXiv:1103.5870 [astro-ph.CO]].
\bibitem{Wetterich:1987fmyRatra:1987rm} 
  C.~Wetterich,
  Nucl.\ Phys.\ B {\bf 302}, 668 (1988);  B.~Ratra and P.~J.~E.~Peebles,
  Phys.\ Rev.\ D {\bf 37}, 3406 (1988).
\bibitem{Peebles:2002gy} 
  P.~J.~E.~Peebles and B.~Ratra,
  Rev.\ Mod.\ Phys.\  {\bf 75}, 559 (2003)
  [astro-ph/0207347].
\bibitem{Kolb:2005da} 
  E.~W.~Kolb, S.~Matarrese and A.~Riotto,
  New J.\ Phys.\  {\bf 8}, 322 (2006)
  [astro-ph/0506534].

\bibitem{Ellis_Wetterich_Amendola_Gasperini} 
  J.~R.~Ellis, S.~Kalara, K.~A.~Olive and C.~Wetterich,
  Phys.\ Lett.\ B {\bf 228}, 264 (1989); C.~Wetterich,
  Astron.\ Astrophys.\  {\bf 301}, 321 (1995)
  [hep-th/9408025];  L.~Amendola,
  Phys.\ Rev.\ D {\bf 62}, 043511 (2000)
  [astro-ph/9908023];  M.~Gasperini, F.~Piazza and G.~Veneziano,
  Phys.\ Rev.\ D {\bf 65}, 023508 (2002)
  [gr-qc/0108016].

\bibitem{interacc} 
  J.~A.~Casas, J.~Garcia-Bellido and M.~Quiros,
  Class.\ Quant.\ Grav.\  {\bf 9}, 1371 (1992)
  [hep-ph/9204213];
  G.~W.~Anderson and S.~M.~Carroll,
  In *Ambleside 1997, Particle physics and the early universe* 227-229
  [astro-ph/9711288].
  N.~Bartolo and M.~Pietroni,
  Phys.\ Rev.\ D {\bf 61}, 023518 (2000)
  [hep-ph/9908521];
  M.~Pietroni,
  Phys.\ Rev.\ D {\bf 67}, 103523 (2003)
  [hep-ph/0203085];
  L.~P.~Chimento, A.~S.~Jakubi, D.~Pavon and W.~Zimdahl,
  Phys.\ Rev.\ D {\bf 67}, 083513 (2003)
  [astro-ph/0303145];
  L.~P.~Chimento, M.~I.~Forte and G.~M.~Kremer,
  Gen.\ Rel.\ Grav.\  {\bf 41}, 1125 (2009)
  [arXiv:0711.2646 [astro-ph]];
  L.~P.~Chimento, M.~I.~Forte and M.~G.~Richarte,
  Mod.\ Phys.\ Lett.\ A {\bf 28}, 1250235 (2013)
  [arXiv:1106.0781 [astro-ph.CO]];
  C.~S.~Rhodes, C.~van de Bruck, P.~.Brax and A.~C.~Davis,
  Phys.\ Rev.\ D {\bf 68}, 083511 (2003)
  [astro-ph/0306343];
  G.~R.~Farrar and P.~J.~E.~Peebles,
  Astrophys.\ J.\  {\bf 604}, 1 (2004)
  [astro-ph/0307316];
  A.~Gromov, Y.~.Baryshev and P.~Teerikorpi,
  Astron.\ Astrophys.\  {\bf 415}, 813 (2004)
  [astro-ph/0209458];
  R.~Mainini and S.~A.~Bonometto,
  Phys.\ Rev.\ Lett.\  {\bf 93}, 121301 (2004)
  [astro-ph/0406114].

\bibitem{Forte:2015oma} 
  M.~Forte,
  Eur.\ Phys.\ J.\ C {\bf 76}, no. 1, 42 (2016)
  doi:10.1140/epjc/s10052-016-3882-6
  [arXiv:1507.03658 [gr-qc]].

\bibitem{Chen:2011cy} 
  X.~m.~Chen, Y.~Gong, E.~N.~Saridakis and Y.~Gong,
  Int.\ J.\ Theor.\ Phys.\  {\bf 53}, 469 (2014)
  doi:10.1007/s10773-013-1831-9
  [arXiv:1111.6743 [astro-ph.CO]].
\bibitem{Bertolami:2007zm} 
  O.~Bertolami, F.~Gil Pedro and M.~Le Delliou,
  Phys.\ Lett.\ B {\bf 654}, 165 (2007)
  doi:10.1016/j.physletb.2007.08.046
  [astro-ph/0703462 [ASTRO-PH]].
\bibitem{Bertolami:2007tq} 
  O.~Bertolami, F.~G.~Pedro and M.~Le Delliou,
  Gen.\ Rel.\ Grav.\  {\bf 41}, 2839 (2009)
  doi:10.1007/s10714-009-0810-1
  [arXiv:0705.3118 [astro-ph]].

\bibitem{Abdalla:2007rd} 
  E.~Abdalla, L.~R.~W.~Abramo, L.~Sodre, Jr. and B.~Wang,
  Phys.\ Lett.\ B {\bf 673}, 107 (2009)
  doi:10.1016/j.physletb.2009.02.008
  [arXiv:0710.1198 [astro-ph]].

\bibitem{He:2009pd} 
  J.~H.~He, B.~Wang and P.~Zhang,
  Phys.\ Rev.\ D {\bf 80}, 063530 (2009)
  doi:10.1103/PhysRevD.80.063530
  [arXiv:0906.0677 [gr-qc]].

\bibitem{Guo:2007zk} 
  Z.~K.~Guo, N.~Ohta and S.~Tsujikawa,
  Phys.\ Rev.\ D {\bf 76}, 023508 (2007)
  doi:10.1103/PhysRevD.76.023508
  [astro-ph/0702015 [ASTRO-PH]]

\bibitem{Quartin:2008px} 
  M.~Quartin, M.~O.~Calvao, S.~E.~Joras, R.~R.~R.~Reis and I.~Waga,
  JCAP {\bf 0805}, 007 (2008)
  doi:10.1088/1475-7516/2008/05/007
  [arXiv:0802.0546 [astro-ph]].

\bibitem{Pavon:2007gt} 
  D.~Pavon and B.~Wang,
  Gen.\ Rel.\ Grav.\  {\bf 41}, 1 (2009)
  doi:10.1007/s10714-008-0656-y
  [arXiv:0712.0565 [gr-qc]].



\bibitem{Forte:2013fua} 
  M.~Forte,
  Gen.\ Rel.\ Grav.\  {\bf 46}, no. 10, 1811 (2014)
  doi:10.1007/s10714-014-1811-2
  [arXiv:1311.3921 [gr-qc]].

\bibitem{Nojiri:2005pu} 
  S.~Nojiri and S.~D.~Odintsov,
  Gen.\ Rel.\ Grav.\  {\bf 38}, 1285 (2006)
  doi:10.1007/s10714-006-0301-6
  [hep-th/0506212].
\bibitem{Li:2004rb} 
  M.~Li,
  Phys.\ Lett.\ B {\bf 603}, 1 (2004)
  doi:10.1016/j.physletb.2004.10.014
  [hep-th/0403127].


\bibitem{Dunajski:2008tg} 
  M.~Dunajski and G.~Gibbons,
  Class.\ Quant.\ Grav.\  {\bf 25}, 235012 (2008)
  doi:10.1088/0264-9381/25/23/235012
  [arXiv:0807.0207 [gr-qc]].

\bibitem{Wang:2005jx} 
  B.~Wang, Y.~g.~Gong and E.~Abdalla,
  Phys.\ Lett.\ B {\bf 624}, 141 (2005)
  doi:10.1016/j.physletb.2005.08.008
  [hep-th/0506069].

\bibitem{Forte:2016ben} 
  M.~Forte,
  Eur.\ Phys.\ J.\ C {\bf 76}, no. 12, 707 (2016)
  doi:10.1140/epjc/s10052-016-4572-0
  [arXiv:1605.06140 [gr-qc]].

\bibitem{Bamba:2012cp} 
  K.~Bamba, S.~Capozziello, S.~Nojiri and S.~D.~Odintsov,
  Astrophys.\ Space Sci.\  {\bf 342}, 155 (2012)
  doi:10.1007/s10509-012-1181-8
  [arXiv:1205.3421 [gr-qc]].

\bibitem{Chimento:2011dw} 
  L.~P.~Chimento, M.~I.~Forte and M.~G.~Richarte,
  Mod.\ Phys.\ Lett.\ A {\bf 28}, 1250235 (2013)
  doi:10.1142/S0217732312502355
  [arXiv:1106.0781 [astro-ph.CO]].



\bibitem{Chimento:2013se} 
  L.~P.~Chimento, M.~Forte and M.~G.~Richarte,
  Eur.\ Phys.\ J.\ C {\bf 73}, no. 1, 2285 (2013)
  doi:10.1140/epjc/s10052-013-2285-1
  [arXiv:1301.2737 [gr-qc]].

\bibitem{Chimento:2013ira} 
  L.~P.~Chimento and M.~G.~Richarte,
  Eur.\ Phys.\ J.\ C {\bf 73}, 2497 (2013)
  doi:10.1140/epjc/s10052-013-2497-4
  [arXiv:1308.0860 [gr-qc]].

\bibitem{Forte:2018loa} 
  M.~Forte,
  arXiv:1801.08036 [gr-qc].. 



\bibitem{Chimento:2012aea} 
  L.~P.~Chimento and M.~G.~Richarte,
  Phys.\ Rev.\ D {\bf 86}, 103501 (2012)
  doi:10.1103/PhysRevD.86.103501
  [arXiv:1210.5505 [gr-qc]].

\bibitem{Kim:2005at} 
  H.~Kim, H.~W.~Lee and Y.~S.~Myung,
  Phys.\ Lett.\ B {\bf 632}, 605 (2006)
  doi:10.1016/j.physletb.2005.11.043
  [gr-qc/0509040].


\bibitem{Granda:2008dk} 
  L.~N.~Granda and A.~Oliveros,
  Phys.\ Lett.\ B {\bf 669}, 275 (2008)
  doi:10.1016/j.physletb.2008.10.017
  [arXiv:0810.3149 [gr-qc]].



\bibitem{Chimento:2009hj} 
  L.~P.~Chimento,
  Phys.\ Rev.\ D {\bf 81}, 043525 (2010)
  doi:10.1103/PhysRevD.81.043525
  [arXiv:0911.5687 [astro-ph.CO]].

\bibitem{Chimento:2013qja} 
  L.~P.~Chimento and M.~G.~Richarte,
  Eur.\ Phys.\ J.\ C {\bf 73}, no. 4, 2352 (2013)
  doi:10.1140/epjc/s10052-013-2352-7
  [arXiv:1303.3356 [gr-qc]].







\end{thebibliography}
\end{document}